\documentclass{aastex62}

\received{January 1, 2018}
\revised{January 7, 2018}
\accepted{\today}
\submitjournal{ApJ}

\shorttitle{Iron dust around AGBs}
\shortauthors{Marini et al.}

\begin{document}

\title{Discovery of stars surrounded by iron dust in the LMC}

\correspondingauthor{Ester Marini}
\email{ester.marini@uniroma3.it}

\author[0000-0002-0786-7307]{Ester Marini}
\affil{Dipartimento di Matematica e Fisica, Universit\'a degli Studi di Roma Tre \\
Via della vasca navale 84, 00100, Roma, Italy}
\affiliation{INAF, Osservatorio Astronomico di Roma \\
Via Frascati 33, 00077, Monte Porzio Catone, Italy}

\author{Flavia Dell'Agli}
\affiliation{Instituto de Astrof\'{\i}sica de Canarias (IAC), E-38200 La Laguna, Tenerife, Spain}
\affiliation{Departamento de Astrof\'{\i}sica, Universidad de La Laguna (ULL), E-38206 La Laguna, Tenerife, Spain}

\author{Marcella Di Criscienzo}
\affiliation{INAF, Osservatorio Astronomico di Roma \\
Via Frascati 33, 00077, Monte Porzio Catone, Italy}

\author{Simonetta Puccetti}
\affiliation{ASI, Via del Politecnico, 00133 Roma, Italy}

\author{D. A. Garc\'{\i}a--Hern\'andez}
\affiliation{Instituto de Astrof\'{\i}sica de Canarias (IAC), E-38200 La Laguna, Tenerife, Spain}
\affiliation{Departamento de Astrof\'{\i}sica, Universidad de La Laguna (ULL), E-38206 La Laguna, Tenerife, Spain}

\author{Lars Mattsson}
\affiliation{Nordita, KTH Royal Institute of Technology and Stockholm University \\
Roslagstullsbacken 23, SE-106 91 Stockholm, Sweden}

\author{Paolo Ventura}
\affiliation{INAF, Osservatorio Astronomico di Roma \\
Via Frascati 33, 00077, Monte Porzio Catone, Italy}



\begin{abstract}

We consider a small sample of oxygen-rich, asymptotic giant branch stars in the 
Large Magellanic Cloud, observed by the \emph{Spitzer} Space Telescope, exhibiting a peculiar 
spectral energy distribution, which can be 
hardly explained by the common assumption that dust around AGB stars is primarily composed
of silicate grains. We suggest that this uncommon class of objects are the progeny of
a metal-poor generation of stars, with metallicity $Z \sim 1-2\times 10^{-3}$, formed 
$\sim 100$ Myr ago. The main dust component in the circumstellar envelope is solid iron.
In these stars the poor formation of silicates is set by the strong nucleosynthesis
experienced at the base of the envelope, which provokes a scarcity of magnesium atoms and
water molecules, required to the silicate formation. The importance of the present results
to interpret the data from the incoming \emph{James Webb Space Telescope} is also discussed.

\end{abstract}



\section{Introduction} \label{sec:intro}

The Large Magellanic Cloud (LMC) has been extensively used as laboratory
to test stellar evolution theories and dust formation mechanisms, owing to
its relative proximity \citep[$\sim 50$ Kpc,][]{feast99} and a low average 
reddening \citep[$E(B-V) \sim 0.075$,][]{schlegel98}.

The evolved stellar population of the LMC have been observed by several
surveys, the two most recent and complete being the Two Micron All-Sky Survey 
\citep{skrutskie06} and the Surveying the Agents of a Galaxy's Evolution Survey (SAGE), 
with the \emph{Spitzer} Space Telescope \citep{meixner06}. The availability of this robust
body of observational data has allowed a full exploration of the main properties
of stars evolving through the Asymptotic Giant Branch \citep[hereafter AGB,][]{marigo99, 
marigo03} and to study the dust enrichment from stellar
sources \citep{srinivasan09, riebel12}

The development of updated AGB models, in which the evolution of the star is coupled
to the description of dust formation in their wind \citep{ventura12, ventura14b,
nanni13, nanni14} opened the way to the characterization of the individual
sources detected by the aforementioned surveys \citep{flavia14a, flavia15}
and to estimate the overall dust production rate by AGB stars currently evolving
in the LMC \citep{raffa14, svitlana}.

Among the various observations in the framework of the \emph{Spitzer} space mission, 
the SAGE-Spec survey \citep{kemper10}
obtained with the Infrared Spectrograph (IRS) is particularly important,
since the analysis of the whole spectral energy distribution (SED) offers the opportunity 
of determining the overall
bolometric flux and the mineralogy of the dust present in the circumstellar
envelope \citep{jones14}. The results from IRS have been recently used to deduce
the fluxes of the AGB stars that will be detected by the MIRI camera, mounted onboard of the 
\emph{James Webb Space Telescope} \citep[JWST, e.g.][]{jones17}; the launch of the JWST 
will open the possibility of extending this research to other Local Group galaxies.

In this work we attempt to explain the peculiar SED exhibited by a small sample of 
evolved M-stars in the LMC, which cannot be explained within the commonly assumed
framework, that silicates are the dominant dust species formed in the wind of M stars.

Our analysis is driven by the changes in the surface chemical composition of
AGB stars, particularly the evolution of the surface abundance of the 
chemical species involved in the formation of the main dust compounds produced in the 
wind of M stars. Based on the behaviour of silicon, magnesium and oxygen, we offer an 
innovative interpretation, suggesting that the dust in the circumstellar envelope of these 
objects corresponds to a rather unusual mineralogy, where solid iron is the dominant
dust species.

The formation of dust mainly composed by solid iron particles in the winds of metal-poor,
evolved stars finds support from independent observational evidences;
\citet{iain10, iain11} studied 14 metal-poor ([Fe/H]~=~$-1.91$ dex to $-0.98$ dex) giant stars 
(including AGB stars) in the Galactic globular cluster $\omega$ Centauri and found
that metallic iron dominates dust production in metal-poor, oxygen-rich stars;
\citet{Kemper02} found that metallic Fe is supposedly dominating the near-infrared flux of 
the M star 
OH 127.8+0.0. Taken together, theory and observations seem to create a picture 
where metallic Fe is a natural constituent of AGB winds. Our results in this Letter 
seem to reinforce this picture.

The conditions required to produce such a peculiar dust composition are achieved 
only by stars within a narrow range of ages and metallicities; therefore, the results 
presented here, if confirmed, can be used as an independent identifier of chemical composition and
formation epoch of the sources observed. These findings will be important for the analysis 
of the soon-to-be results from the JWST mission, particularly when studying galaxies 
dominated by a metal poor stellar component.

\begin{figure}
\resizebox{1.\hsize}{!}{\includegraphics{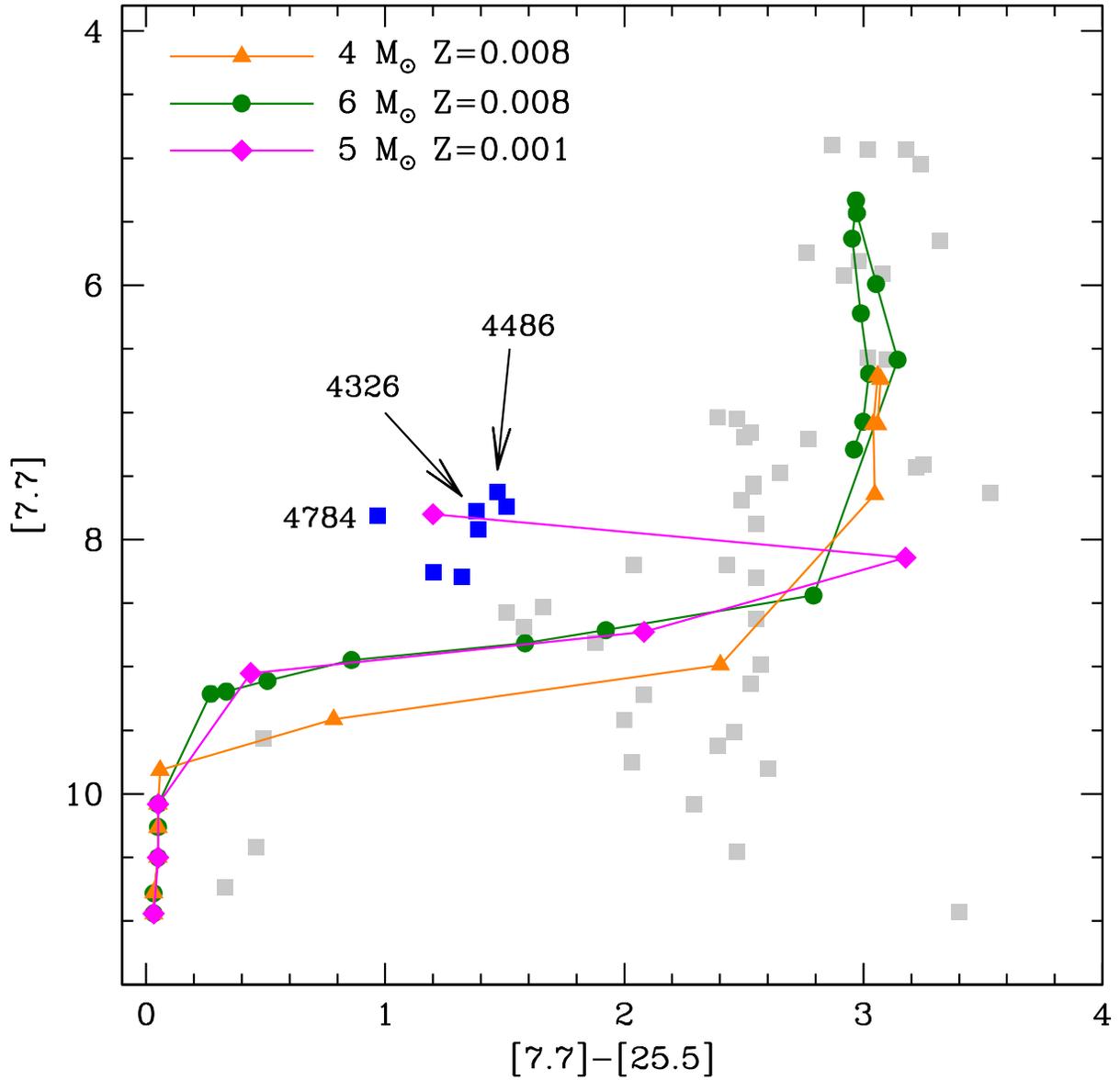}}
\vskip-120pt
\caption{The distribution of the M-stars sample of the LMC analyzed by \citet{jones17},
indicated by grey squares, in the color-magnitude ($[7.7]-[25]$, $[7.7]$) plane. Blue 
squares refer to the group of stars discussed in the present investigation. 
Green points and orange triangles indicate the synthetic colours assumed by stars of 
metallicity $Z=8\times 10^{-3}$ and mass, respectively, $4~M_{\odot}$ and $6~M_{\odot}$, 
during the AGB evolution. Magenta diamonds refer to the evolution of a $5~M_{\odot}$ star
of metallicity $Z=10^{-3}$, before the achievement of the C-star stage.}
\label{fcmd}
\end{figure}

\section{Iron Dusty AGB stars in the LMC}
In a recent work \citet{jones17} analyzed the SED of evolved stars in the LMC observed
within the SAGE-Spec survey and provided the magnitudes expected in the mid-IR bands 
of the MIRI camera of the JWST. 

Fig.~\ref{fcmd} shows the distribution of the M stars analyzed by \citet{jones17} in
the colour-magnitude ($[7.7]-[25]$, $[7.7]$) plane (hereafter CMD). The position
of the stars in this plane allows the 
best discrimination of the mineralogy of the dust present in the circumstellar envelope, 
as the relative distribution of the various dust species mostly
affects the details of the shape of the SED in the region covered
by the MIRI filter centered at $7.7 \mu$m.

\begin{figure*}
\begin{minipage}{0.33\textwidth}
\resizebox{1.\hsize}{!}{\includegraphics{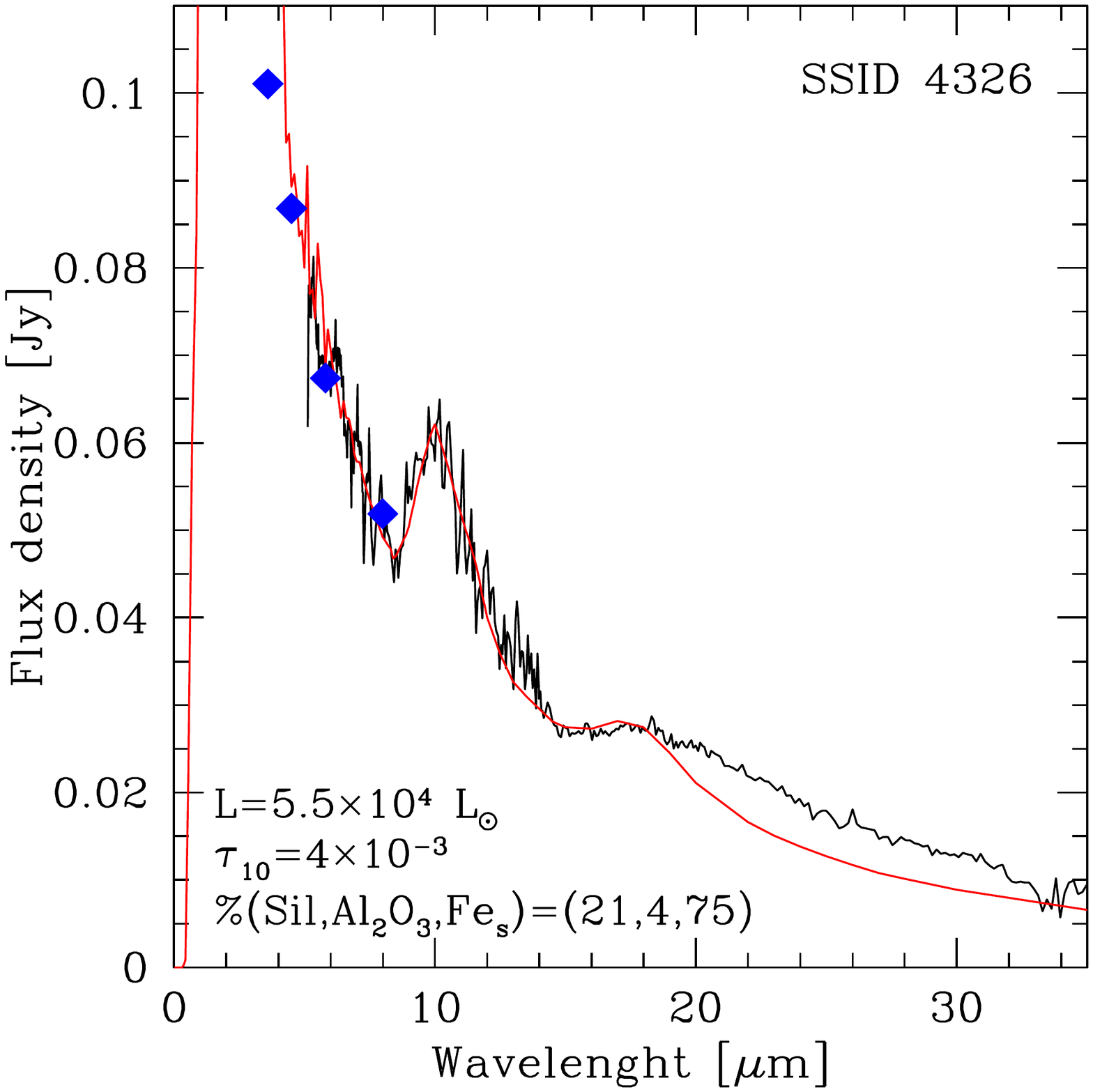}}
\end{minipage}
\begin{minipage}{0.33\textwidth}
\resizebox{1.\hsize}{!}{\includegraphics{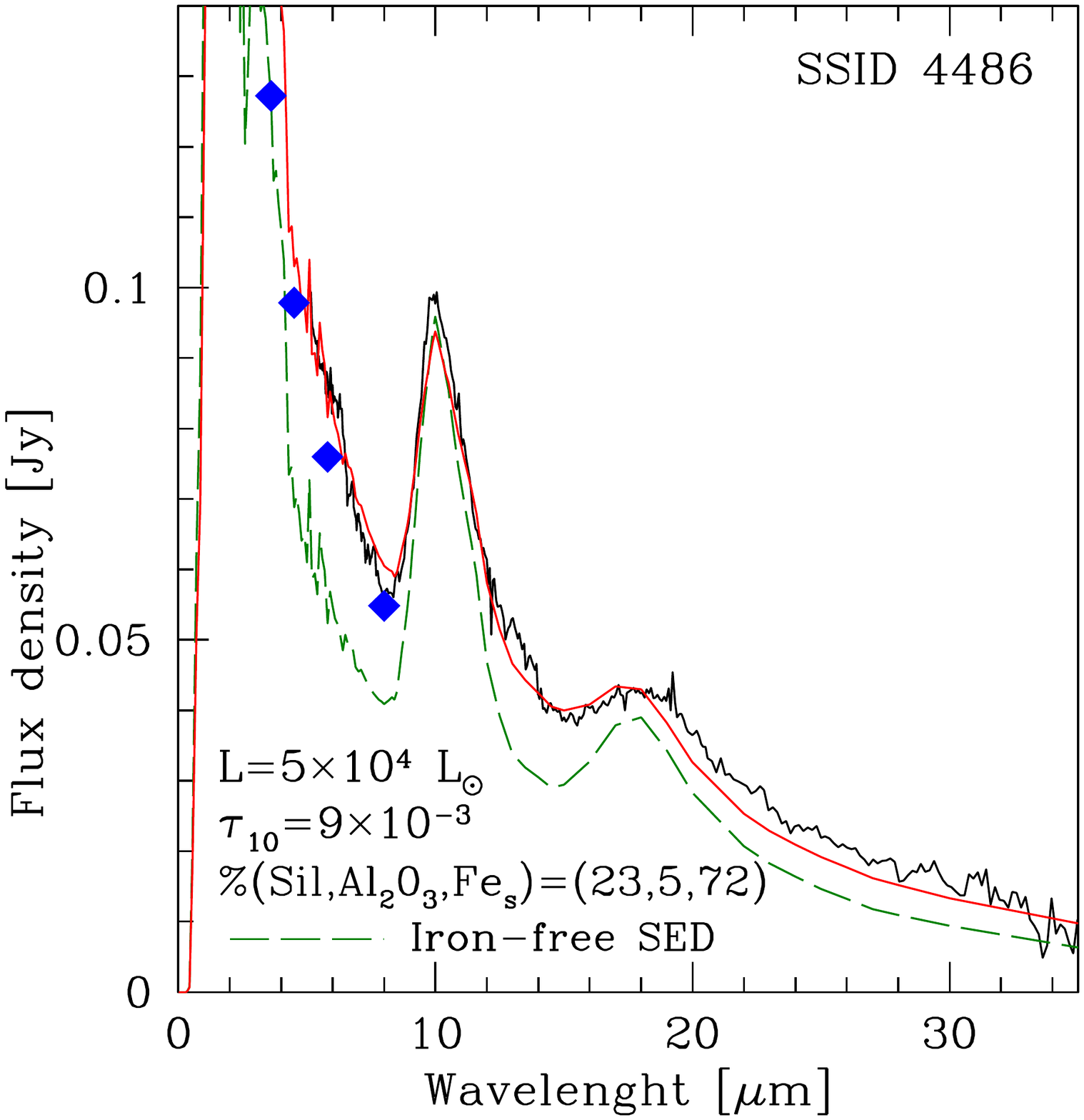}}
\end{minipage}
\begin{minipage}{0.33\textwidth}
\resizebox{1.\hsize}{!}{\includegraphics{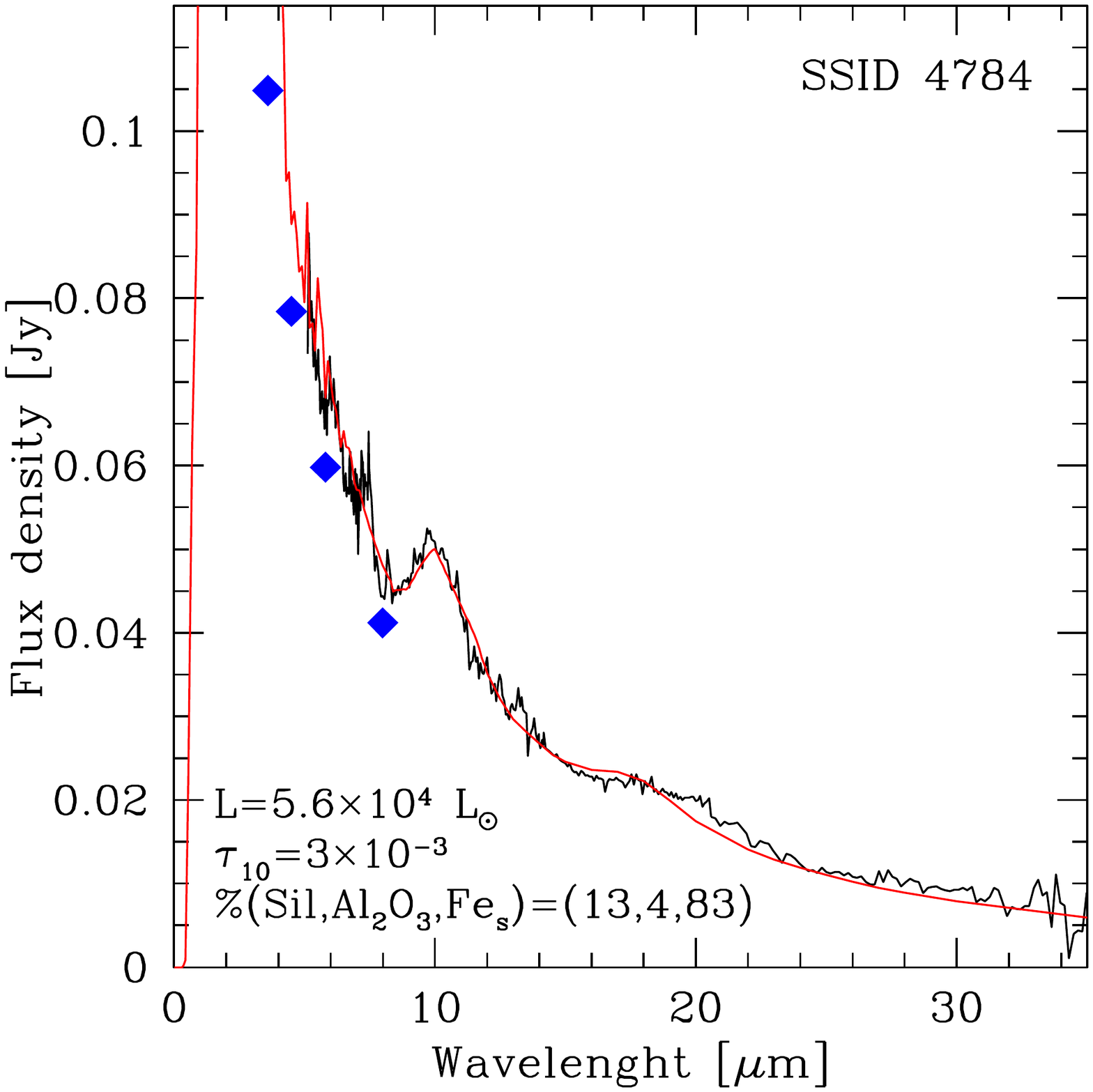}}
\end{minipage}
\vskip-40pt
\caption{The comparison between the observed IRS spectra taken from
\citet{woods11} (in black) and the synthetic SED (in red), obtained by assuming 
different luminosities and percentages of the various dust species, for three stars in the 
LMC, interpreted as low-metallicity massive AGBs (see text for discussion). Blue diamonds
refer to IRAC $[3.6]$, $[4.5]$, $[5.8]$, $[8.0]$ photometry, from \citet{meixner06}.
The green, dashed line in the middle panel refers to a synthetic SED, calculated
assuming iron-free dust around the star.}
\label{fsed}. 
\end{figure*}

Here we focus on the group of stars populating the left region of the CMD,
located at $[7.7]-[25] \sim 1.4$, $[7.7] \sim 7.5-8$, not covered by the 
tracks of $Z=8\times 10^{-3}$ stars. These objects, indicated with blue 
squares in Fig.~\ref{fcmd}, exhibit a very peculiar SED; Fig.~\ref{fsed} displays
three illustrative examples. The $9.7\mu$m feature in the SED indicates the presence of silicate 
type dust. On the other hand, the steep rise of the SED for wavelengths shorter than 
$8\mu$m cannot be reproduced if we assume that most of the dust is composed of silicates.
This is shown in the middle panel of Fig.~\ref{fsed}, where we report the SED obtained
assuming the same luminosity of the best-fit model, an optical depth chosen to
reproduce the morphology of the silicate feature, and iron-free dust: the comparison 
with the observed SED clearly shows that the overall SED, both in the spectral region 
$\lambda < 8\mu$m and at the mid-IR wavelengths $\lambda > 12\mu$m, does not
match the observations.

Before presenting our interpretation on the evolutionary status of these objects, we
briefly recall the most relevant physical and chemical properties of stars evolving
through the AGB phase.

\section{The evolution through the asymptotic giant branch}
The AGB phase is experienced by all the stars of mass in the range 
$0.8~M_{\odot} < M < 8~M_{\odot}$. The pollution from these objects is associated to 
the changes in the surface chemical composition, caused by 
third dredge-up (TDU) and hot bottom burning (HBB). The former provokes a carbon enrichment 
of the surface regions, which can eventually lead to the formation of 
a carbon star \citep{iben74}. HBB is related to the ignition of a p-capture nucleosynthesis 
at the base of the convective envelope, when the temperature ($T_{\rm bce}$) exceeds 
$\sim 30-40$ MK \citep{blocker91}; pollution from HBB reflects the equilibria of the p-capture 
reactions activated in the innermost layers of the envelope, which, in turn, are sensitive 
to $T_{\rm bce}$.

\subsection{Hot bottom burning in massive AGB stars}
The ignition of HBB requires core masses above $\sim 0.8~M_{\odot}$, which reflects into
initial masses $M>3~M_{\odot}$. On general grounds, the higher the mass the stronger the
HBB activated, because stars of higher mass evolve on bigger cores during the AGB phase.
The strength of HBB is extremely sensitive to the metallicity, because lower Z stars
attain hotter $T_{\rm bce}$'s, thus experiencing a more advanced nucleosynthesis at the
base of the envelope \citep{ventura13}. The description of HBB is highly
sensitive to convection modelling \citep{ventura05}, which is the reason for the significant 
differences obtained by the various groups studying AGB evolution.

Fig.~\ref{fmodels} shows the evolution of the luminosity (left panel) and of the 
surface mass fractions of $^{16}$O and $^{24}$Mg (middle panel) of AGB stars of different
mass and metallicity $Z=10^{-3}$. The ignition of HBB is
witnessed by the drop in the surface abundances of oxygen and magnesium, a signature
of the activation of full CNO cycling and of the Mg-Al-Si nucleosynthesis. The strength 
of HBB is sensitive to the mass of the star: 
the $4~M_{\odot}$ star experiences only a soft HBB,
with the depletion of oxygen being limited to a factor of $\sim 2$, occurring only
during the very late AGB phases. Conversely, in $5-6~M_{\odot}$ stars, a significant
depletion of both oxygen and magnesium takes place since the early AGB phases.

\subsection{Dust formation in oxygen-rich AGB stars}
The changes in the surface chemical composition reflects into the dust formed in the
circumstellar envelopes of AGB stars. In the high-mass domain the effects of HBB
provoke the destruction of the surface carbon, leaving no space for the formation of
carbonaceous particles\footnote{This is connected with the high stability of the CO
molecule \citep{sharp90}, which makes all residual carbon to be locked into CO in 
oxygen-rich stars}. Under these conditions the most stable dust species are alumina
dust (Al$_2$O$_3$) and silicates. Solid iron particles also form, but in modest
quantities, because this species is less stable than Al$_2$O$_3$ and silicates, thus
it forms in more external and less dense environments \citep{fg06}. 

Al$_2$O$_3$ forms in a region $\sim 2$ stellar radii away from the photosphere of the 
star \citep{flavia14b}; this very stable compound is extremely transparent to the 
electromagnetic radiation, thus its formation has negligible effects on the dynamics of 
the wind. 

Among silicates, the most relevant species is olivine (Mg$_2$SiO$_4$), which forms when 
the temperature of the gas drops below $\sim 1100$ K, in a region of the circumstellar 
envelope located $\sim 5-7$ stellar radii from the photosphere of the star. Formation of 
olivine particles occurs via a reaction involving SiO and water molecules and magnesium 
atoms; consequently, the amount of Mg$_2$SiO$_4$ which can be formed is constrained by 
the number densities of silicon and magnesium, and by the excess of oxygen with respect 
to carbon and silicon \citep{fg06}. As far as there are magnesium and oxygen available, 
silicates are the most abundant dust species, because the surface content 
of silicon and magnesium are higher than aluminium and iron.

\begin{figure*}
\begin{minipage}{0.33\textwidth}
\resizebox{1.\hsize}{!}{\includegraphics{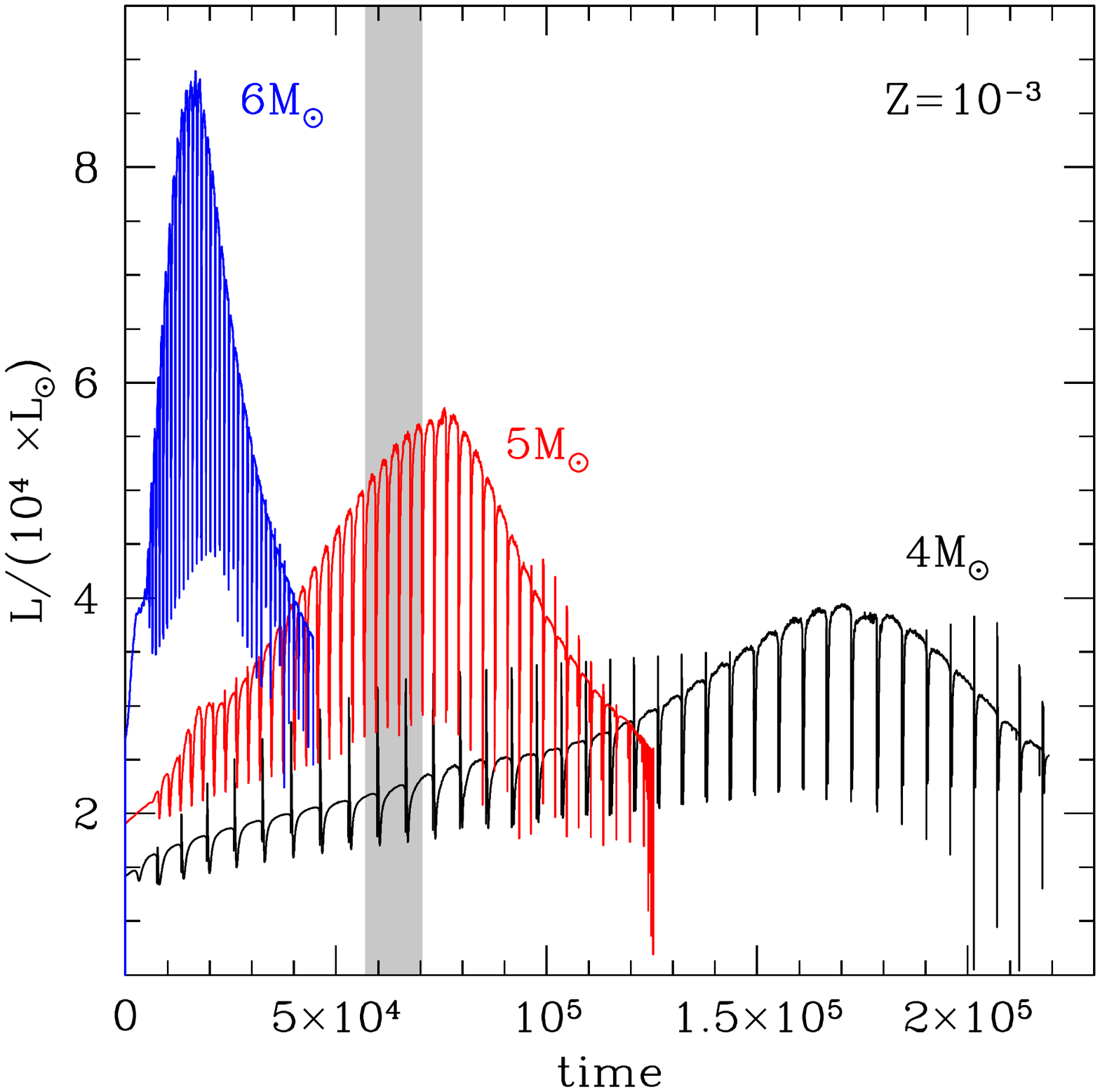}}
\end{minipage}
\begin{minipage}{0.33\textwidth}
\resizebox{1.\hsize}{!}{\includegraphics{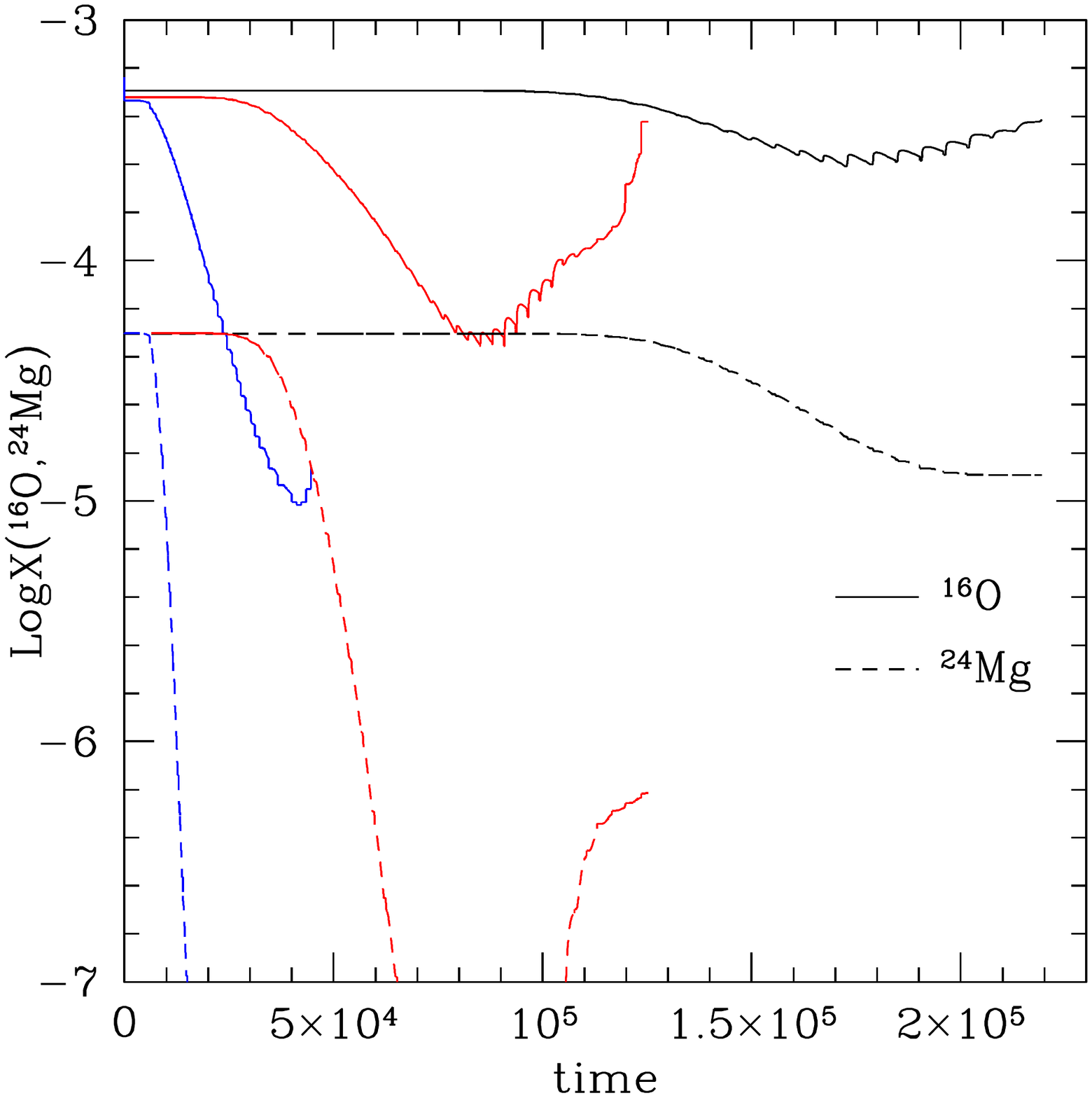}}
\end{minipage}
\begin{minipage}{0.33\textwidth}
\resizebox{1.\hsize}{!}{\includegraphics{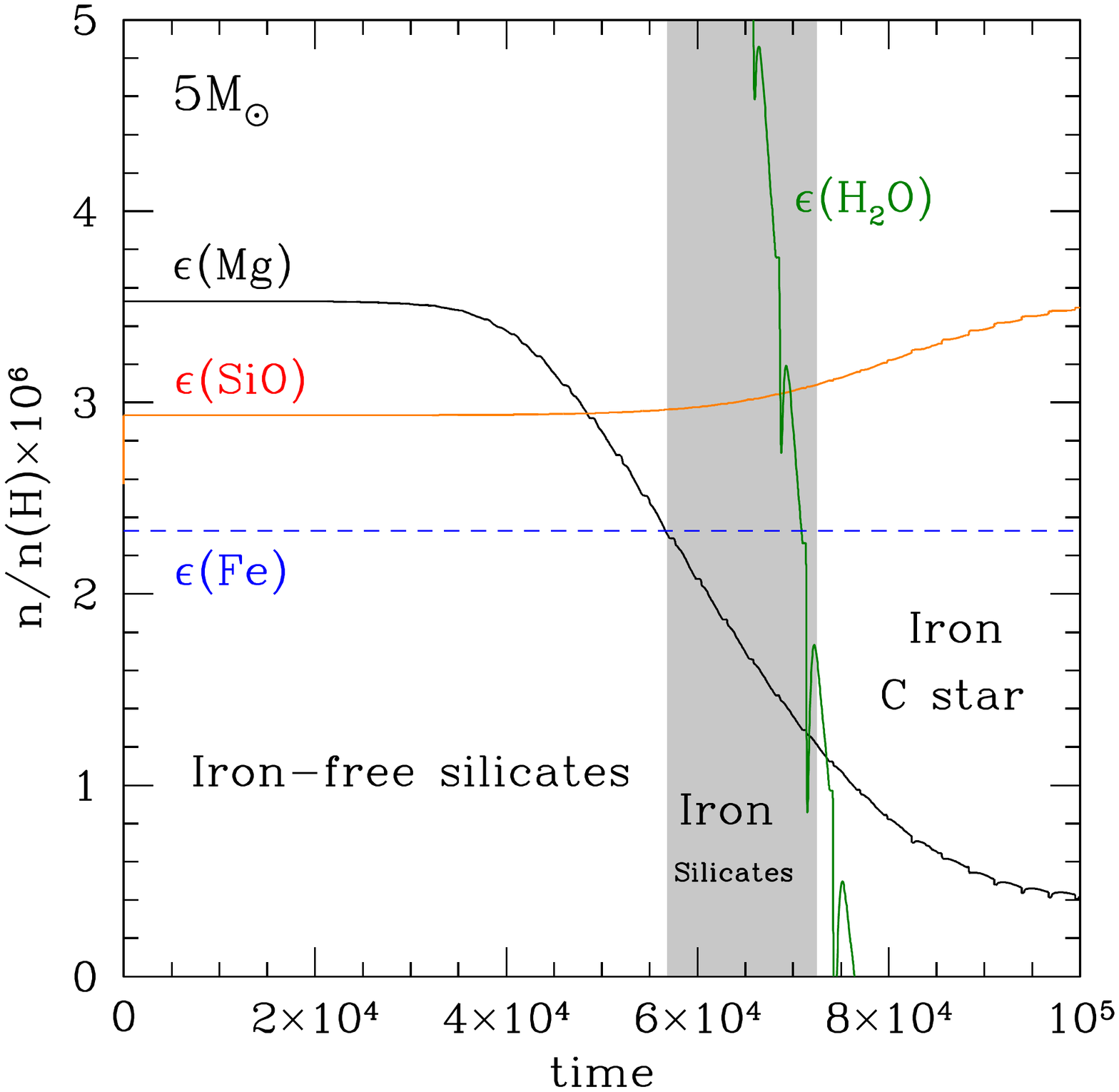}}
\end{minipage}
\vskip-40pt
\caption{The variation of the main physical and chemical properties
of stars of metallicity $Z=10^{-3}$ and initial mass $4~M_{\odot}$ (black lines), 
$5~M_{\odot}$ (red) and $6~M_{\odot}$ (blue), during the AGB phase. Times on the abscissa 
are counted
since the beginning of the thermally pulsating phase. The left and middle panels show, respectively,
the variation of the luminosity and of the surface mass fractions of $^{16}$O (solid)
and $^{24}$Mg (dashed). The right panel, which refers to the $5~M_{\odot}$ model,
shows the AGB variation of the main quantities relevant to dust production, namely 
the number densities of silicon (red), magnesium (black), iron (blue, dashed) and water 
molecules (green), the latter being given by 
the oxygen excess with respect to carbon and silicon; all the four quantities are 
normalized to the number density of hydrogen. We mark the three AGB phases 
during which dust in the circumstellar envelope has the following composition: 
a) mainly silicates; b) a dominant solid iron contribution with traces of silicates and 
alumina dust; c) no silicates formation is possibile, due to lack of water 
molecules. Grey shading in the left and right panels indicate the AGB phases of the
$5~M_{\odot}$ star corresponding to phase (b).}
\label{fmodels}
\end{figure*}

The right panel of Fig.~\ref{fmodels} shows the variation of the three 
aforementioned
number densities and iron during the AGB evolution of the $5~M_{\odot}$ star. All the 
quantities are normalized to the hydrogen density.
 
We can distinguish 4 phases. 

\begin{enumerate}

\item{
In the initial AGB phases the surface magnesium is in 
excess of silicon, thus the rate of formation of silicates is constrained by the
silicon abundance; in this regime the least abundant among the species involved in the
formation reaction of olivine are the SiO molecules.}

\item{ 
The ignition of HBB provokes 
the depletion of the surface Mg, thus the key species for the formation of
silicates is magnesium; the dust formed is still dominated by silicates.}

\item{ 
The action of HBB eventually makes the Mg$/$Fe ratio to drop below unity; in these
conditions the formation of silicates is severely reduced, thus leaving
room for the formation of solid iron dust.}

\item{ 
During the final AGB phases the surface oxygen mass fraction becomes so small that 
the star becomes a C-star. The surface C is so low that iron dust is expected to be the
dominant species anyway.}

\end{enumerate}

These results show that dust production in low-Z, massive AGB stars can deviate from the
common assumption that silicates is the dominant species, rather indicating that
during the advanced AGB phases, after the effects of HBB have severely modified the
surface chemical composition, there is wide room to the formation of iron grains.

Formation of metallic Fe in AGB atmospheres seems plausible from the expected 
condensation temperature (similar to that of silicates) at gas pressures typical of 
circumstellar environments \citep{Gail99,fg06}. Laboratory experiments have
shown that metallic Fe condenses almost ideally under high supersaturation conditions, 
and evaporates nearly ideally in vacuum \citep{Tachibana11}. However, metallic Fe does 
not necessarily nucleate homogeneously in circumstellar environments, but likely through 
heterogeneous nucleation on pre-existing dust, e,g., the highly stable Al$_2$O$_3$. 
For a supersaturation ratio $S>1$, the condensation and evaporation efficiencies are almost 
equal \citep{Tachibana11}, which means that $S >1$ is in principle sufficient to grow metallic Fe. Thus, if 
silicate formation is suppressed due to low Mg abundance, metallic Fe can form and grow since 
efficient evaporation is essentially prevented in the presence of metallic iron vapour.

\section{Discussion}
In Fig.~\ref{fcmd} we show two evolutionary AGB sequences, corresponding to
stars of initial masses $4~M_{\odot}$ and $6~M_{\odot}$, with $Z=8\times 10^{-3}$
\citep{flavia15};
this is the metallicity shared by the majority of the stars in the LMC younger
than $200$ Myr \citep{harris}.
The evolutionary paths in the figure can be explained
by the significant production of silicates which starts after the beginning of 
HBB \citep{ventura12, flavia15}. The tracks first move to the red, almost horizontally, owing 
to the gradual rise of the mid-IR flux as the circumstellar envelope becomes more 
and more opaque. The progressive increase in the rate of dust production eventually favours 
the formation of a prominent silicate emission feature at $9.7\mu$m, which lifts the flux 
in the $7.7 \mu$m region, thus provoking a vertical upturn in the tracks.

Fig.~\ref{fcmd} shows that the position of most M stars discussed by \citet{jones17}
is reproduced by the evolutionary tracks: therefore, they
can be explained within the traditional understanding, that
dust formation around O-rich AGB stars is mainly composed by silicates, with
little traces of alumina dust. 

On the other hand, the colours and magnitudes of the peculiar stars, indicated with blue 
squares in Fig.~\ref{fcmd}, are not reproduced by the $Z=8\times 10^{-3}$ tracks. As
discussed in section 2 and shown in Fig.~\ref{fsed}, their SED cannot be reproduced
if we assume that silicates are the main dust species.
We propose that the dust formed in the wind of these peculiar objects is mainly composed
by iron grains. Indeed, Fig.~\ref{fsed} shows that a highly satisfactory 
fit of the observed SED is obtained by assuming $\sim 70-80\%$ of solid iron, with smaller 
percentages of silicates ($\sim 15-20\%$) and alumina dust ($\sim 5\%$). We suggest that 
these stars descend from metal-poor progenitors with initial masses $\sim 5-6~M_{\odot}$,
formed $\sim 100$ Myr ago. The presence of a recent low-metallicity star formation activity  
is compatible with the star formation history and the age-metallicity evolution of the LMC 
computed by \citet{harris}. The particular mineralogy of the 
dust in the envelope of these objects is determined by the action of HBB, which, as 
described in the previous section, partly inhibits the production of silicates. 

The evolutionary track of the $5~M_{\odot}$ star discussed in Fig.~\ref{fmodels} is
shown in Fig.~\ref{fcmd}. During the first part of the evolution, corresponding to
the phases (i) and (ii) above, the path followed is similar to the higher
metallicity, massive AGBs: the track moves to the red, owing to the
formation of silicates in the circumstellar envelope. However, when iron takes over
as the dominant dust species (point iii above, grey-shaded region in Fig.~\ref{fmodels}) 
the track moves to the left, 
entering the region populated by the stars analyzed in the present investigation.
According to these results, all the stars in the regions of the CMD at $[7.7] < 7$ 
are massive AGBs of metallicity $Z \geq 4\times 10^{-3}$.

Only low-metallicity stars of mass $M > 4~M_{\odot}$ evolve through a phase dominated
by iron dust, 
because the HBB experienced by lower mass objects is not sufficiently strong to trigger
a significant decrease in the surface Mg and O (see middle panel of Fig.~\ref{fmodels}).
The constraint on the initial mass is fully consistent with the luminosities required to 
reproduce the  overall SED, of the order of $50.000L_{\odot}$ (see grey-shaded areas
in the left and right panels of Fig.~\ref{fmodels}). This interpretation is confirmed by 
the observed periods of the sources discussed, which are in the range $600-640$ d
\citep{fraser08, martin18}: these are the same periods
expected for the $5~M_{\odot}$ star when evolving through the phase dominated
by iron dust (grey-shaded regions in Fig.~\ref{fmodels}), calculated by applying eq.~4 in 
\citet{vw93}.  

We rule out that the metallicity of the sources considered here 
is above $Z > 2\times 10^{-3}$, because the HBB experienced is not sufficiently strong to
allow a significant O and Mg depletion in the surface regions \citep{ventura14a}. 
$Z \sim 10^{-3}$ represents the lowest metallicity that dust around M stars 
is currently observed to be forming at \citep{boyer17}.

The stars discussed here populate a specific region in the CMD; this opens the 
way to the definition of a criterion to identify metal-poor stars exposed to HBB in samples
of AGB sources, which will be extremely important to 
interpret the data from the JWST space mission.

The inhibition of silicate production, which is related to the lack
of magnesium, confirms that very low-metallicity, massive AGB stars experience strong HBB
at the base of their envelope, thus suggesting that convection is extremely efficient
in those physical conditions, far in excess of what is required to reproduce the evolution 
of the Sun.

\section{Conclusions}
We study a peculiar class of M-type AGB stars in the LMC, whose spectral energy distribution 
exhibits a peculiar shape, which cannot be explained by invoking a dominant contribution 
from silicates to the dust in the circumstellar envelope.

We propose that these sources descend from $\sim 5~M_{\odot}$ stars, $\sim 100$ Myr old, 
of metallicity $Z \sim 1-2\times 10^{-3}$. The peculiar SED is due to the uncommon 
mineralogy of the dust in the circumstellar envelope, which is mainly ($\sim 80\%$) 
composed by solid iron, completed by silicates and alumina dust particles. This 
distribution is favoured by the effects of hot bottom burning, which provokes a shortage 
of magnesium and water molecules, two essential ingredients for the formation of silicates.

Theoretical arguments show that gaseous iron is expected to condense efficiently 
in metallic iron in the winds of AGB stars; furthermore, there are 
observational evidences that iron grains are an important opacity source at low 
metallicities. However, this has been remained undetected for 
long, due to the iron featureless spectrum. In this work we have demonstrated the 
possibility of recognizing the presence of such iron grains from the observed SED of 
AGB stars. 

This findings opens the way to use the
incoming JWST observations to identify metal-poor, young stars, 
because these peculiar objects are expected to
populate well identified regions in some of the colour - magnitude planes built with 
the MIRI filters. Side results of this study is the evidence that metal-poor,
massive AGBs experience an advanced p-capture nucleosynthesis at the base of the
envelope, a result still highly debated \citep[see e.g.][]{karakas14}.

\acknowledgments
EM and PV are indebted to Franca D'Antona for helpful discussions.
FDA and DAGH acknowledge support provided by the Spanish Ministry of Economy and 
Competitiveness (MINECO) under grant AYA-2017-88254-P.



\end{document}